\documentclass[%
 reprint,
 amsmath,amssymb,
 aps,superscriptaddress,
]{revtex4-2}

\usepackage{xcolor}
\usepackage{graphicx}
\usepackage{dcolumn}
\usepackage{bm}

\begin{document}

\preprint{APS/123-QED}

\title{Characterization of the inner edge of the neutron star crust}

\author{R. Shafieepour}
\address{Department of Physics, University of Tehran, Tehran 14395-547, Iran}
\author{H. R. Moshfegh}
\address{Department of Physics, University of Tehran, Tehran 14395-547, Iran}
\author{J. Piekarewicz}
\address{Department of Physics, Florida State University, Tallahassee, Florida 32306-4350, USA}

\date{January 1, 2022}

\begin{abstract}
The poorly known crustal equation of state plays a critical role in many observational phenomena associated with a neutron star. 
Using semi-classical Monte Carlo simulations, we explore the possible configurations of the inner edge of the neutron-star crust 
for a variety of baryon densities and proton fractions. Applying the Kirkwood--Buff theory to these two-component systems, we
observe how the isothermal compressibility reaches a maximum when isolated non-symmetric clusters are formed in an
extremely dilute neutron gas. To determine the neutron fraction, we suggest a 
geometrical model based on the behavior of the proton-neutron pair correlation function. Accordingly, the equation of state of the 
inner crust is calculated, illustrating that the nuclear energy in $\beta$-equilibrium follows a power-law behavior with baryon density. 
As a possible astrophysical outcome of this study, our results could help refine the mass--radius relation. Finally, our results pave 
the way towards further investigations of the impact of the proton-neutron pair correlation function on transport properties within 
the neutron-star crust.

\end{abstract}

\maketitle
\section{INTRODUCTION}
Neutron stars, with maximum masses of about $2M_{\odot}$ \cite{R1,R2,R3}, are unique astronomical objects 
that probe densities that are inaccessible in terrestrial laboratories. Given that neutron stars are assumed to be 
in hydrostatic equilibrium, their density ranges from a few times nuclear saturation ($\rho_0$ = 0.16 fm$^{-3}$) 
all the way down to terrestrial densities \cite{R4}. Although bound by gravity and not by the strong force, neutron 
stars may be regarded as large spherical nuclei with a density gradient in the radial direction \cite{R5}. It has been 
argued that the stellar core is comprised of a uniform neutron-rich liquid surrounded by a non-uniform crust composed
of exotic non-spherical clusters often referred to as “nuclear pasta” phases \cite{R6}. These pasta phases emerge 
from a competition between the long-ranged Coulomb repulsion and the short-ranged nuclear attraction. At baryon
densities below $\rho_0$ \cite{R7}, these length scales are comparable leading to a universal behavior known as
Coulomb frustration. Besides their exotic nature, it has also been shown that the transport properties of the inner 
crust strongly depend on the geometry of the pasta phases \cite{R8,R9}. Further, the presence of pasta phases 
could explain some observations such as a delay in neutrino signals \cite{R10,R11}, the lack of X-ray-emitting 
isolated pulsars with long spin periods ($>\!12$\,s) \cite{R12}, and the slow cooling rate of MXB 1659-29 \cite{R13}.

Molecular dynamics (MD) and Monte Carlo (MC) simulations provide a suitable platform for studying the emergence
of the various pasta phases without any biases on the initial shapes \cite{R14,R15,R16,R17}. Notably, these 
semi-classical simulations can accommodate a very large number of particles, resulting in minimizing the finite size effect 
that hinders other approaches. Recently, Schneider \emph{et al.,} have studied the formation of multiple domains and 
defects in the pasta structures using a simulation volume containing about three million nucleons \cite{R18}. As 
Coulomb frustration is characterized by the development of a large number of nearly degenerate ground states, 
small changes in certain thermodynamic properties --- such as baryon density, proton fraction, and temperature ---
have a dramatic impact on the topology of the pasta phases and ultimately on their transport properties. Indeed,
using quantum MD simulations at zero temperature, Nandi and Schramm have shown that the spherical clusters 
with well-defined surfaces transform into elongated spaghetti-like shapes by increasing the baryon density from 
about 0.017 to 0.034 fm$^{-3}$ \cite{R19}. Further, it has been illustrated that the nuclear waffle phase, consisting
of a stack of perforated parallel plates, quickly merged by increasing the temperature from 1 to 1.30 MeV \cite{R20}. 
In contrast, as the system got cooled to 0.75 MeV, the position and shapes of the holes ceased to fluctuate, ultimately
forming a two-dimensional hexagonal lattice \cite{R20}. In turn, Piekarewicz and Toledo-S\'anchez have shown that in
$\beta$-equilibrium, the electronic contribution --- required to maintain overall charge neutrality --- dominates the 
total energy of the system, leading to very small proton fractions at all densities of relevance to the inner crust 
\cite{R21}. Using a lattice-gas model with nearest-neighbors attractive interactions together with a long-range Coulomb 
repulsion, the charged Ising model revealed how the heat capacity reaches a maximum as nuclear clusters
coexist with a dilute neutron vapor at a temperature of 3 MeV and for a baryon density and proton fraction 
of $\sim$0.03 fm$^{-3}$ and 0.3, respectively \cite{R22}. MD simulations have also been performed to study 
transport properties in the stellar crust. Horowitz and Berry have shown that the shear viscosity of the nuclear 
pasta increases slightly relative to the lower density phase that is dominated by spherical clusters --- in contrast 
to the conventional behavior in complex fluids where the shear viscosity depends dramatically on the shape of 
the molecules \cite{R23}. Following a similar approach, Caplan, Forsman, and Schneider used “lasagna” and
“disordered” configurations to show that the shear viscosity reaches its minimum at the melting temperature
\cite{R24}. Finally, Kycia and collaborators have shown that algebraic methods that invoke the Euler characteristic 
and Betti numbers associated with the different shapes can shed light on the finer topology of pasta and sub-pasta 
phases \cite{R25}.

In the present work we perform MC simulations for baryon densities in the range 0.005--0.08 fm$^{-3}$ (or 
from about 0.03 to 0.5\,$\rho_0$) and proton fractions from 0 to 0.4, albeit for a relatively modest number 
of nucleons ($A\!=\!5000$). Although the number of nucleons is modest, the main virtue of the present work
is the use of the Kirkwood--Buff theory \cite{R27} to compute the isothermal compressibility and density 
fluctuations. To our knowledge, no previous attempt has been made to investigate the presence of possible 
critical points in the inner crust of neutron stars. Therefore, the main goal of this manuscript is to identify the
maximum of the isothermal compressibility as a function of the baryon density for different proton fractions. 
In addition, we suggest a geometrical model based on the behavior of the proton--neutron pair correlation 
function to estimate the bound neutron fraction. Isolating the bound and free neutron fractions is relevant 
because free neutrons contribute to the pressure support against gravitational collapse. That is, in addition
to providing a clearer picture of the exotic pasta phases, the free neutron fraction modifies the crustal equation
of state (EoS) which in turn may affect the mass--radius relation \cite{R6,R28}. 

The paper is organized as follows. Sec. II describes the semi-classical MC simulation, the underpinning theoretical 
formalism, and the geometrical model used to estimate the bound neutron fraction. Sec. III is devoted to our results,
which focus on the calculation of the isothermal compressibility, the bound neutron fraction, and its impact on the
equation of state. Finally, we offer our conclusions in Sec. IV.
\section{FORMALISM}

In this section, we describe the theoretical formalism that will be used to simulate nuclear matter in a region
of relevance to the inner crust. We start with a brief description of the Monte Carlo simulation and devote the
rest of the section to computing some relevant neutron-star observables.

\subsection{Semi-classical Monte Carlo simulation}
The MC simulation used in this work follows closely the approach presented in Ref. \cite{R21}. In the interest
of clarity, we provide a concise review of the most important points. The simulation box consists of $A\!=\!5000$ 
nucleons that are randomly distributed initially throughout the box. To simulate baryon densities in the 
0.08 to 0.005 fm$^{-3}$ range, box lengths ($\mathfrak{L}$) ranging from 39.69 to 100 fm were selected together with
proton fractions $y\!=\!Z/A$ varying from 0 (pure neutron matter) to 0.4 (nearly symmetric nuclear matter). 
Nucleons were embedded in a uniform electron background at a density of
$\rho_e\!=\!Z/\mathfrak{L}^{3}$ to enforce charge neutrality. Given that in this density domain relevant to the stellar crust where the kinetic energy dominates, 
the electrons may be treated as a non-interacting Fermi gas. The total potential energy of the $A$-body system 
consists of nuclear and Coulomb interactions,
\begin{equation}
  V({\bf r}_{1},\ldots,{\bf r}_{A})\!=\!V_{\rm Nucl}({\bf r}_{1},\ldots,{\bf r}_{A})+
                                                      V_{\rm Coul}({\bf r}_{1},\ldots,{\bf r}_{A}),
\label{eq:E1}
\end{equation}
where the nuclear interaction is modeled as a sum of two-body terms of the following form:
\begin{equation}
 V_{\rm Nucl}({\bf r}_{1},\ldots,{\bf r}_{A})\!=\!\sum_{i<j=1}^{A}
   \Big[ae^{-r_{ij}^2/\Lambda}+(b+c\tau_i\tau_j)e^{-r_{ij}^2/2\Lambda}\Big],
 \label{eq:E2}
\end{equation}
where $r_{ij}\!=\!|{\bf r}_{i}\!-\!{\bf r}_{j}|$ denotes the inter-particle separation and $\tau$ is the 
isospin of the nucleon: $\tau\!=\!+1$ for a proton and $\tau\!=\!-1$ for a neutron. Although simple, 
the two-body interaction encodes the characteristic nucleon-nucleon short-range repulsion and 
the intermediate-range attraction between proton-neutron pairs. The model parameters are given
by: $a\!=\!110$\,MeV, $b\!=\!-26$\,MeV, $c\!=\!24$\,MeV, and $\Lambda\!=\!1.25\,{\rm fm}^{2}$. 
The isospin dependence of the two-body potential ensures that while pure neutron matter is 
unbound, symmetric nuclear matter saturates at the correct density. Further details of the 
calibration procedure may be found in Ref.\cite{R15}. 

Whereas for the short-range nuclear interaction finite-size effects are minimized by adopting
periodic boundary conditions and implementing the so-called minimum-image convention, such
an approach is not applicable in the case of the long-range Coulomb interaction. Hence, one 
can either introduce a screening length as a result of the polarization of the electron gas \cite{R15}
or can implement an exact Ewald summation \cite{R29}. The great merit of the century-old Ewald 
summation is that the Coulomb contribution can be divided into a short-range part that can be
readily evaluated in configuration space and periodic long-range contribution that converges
rapidly in momentum space. Using the Ewald summation for a system of $Z$ protons immersed in a neutralizing and uniform 
electron background, the Coulomb potential may be exactly calculated.
That is,
\begin{equation}
 V_{\rm Coul}({\bf r}_{1},\ldots,{\bf r}_{A})\!=\! 
   \frac{q^{2}}{\mathfrak{L}}\left[U_{0}+\frac{1}{2}\sum_{i\ne j}\Big(u_{\rm sr}({\bf s}_{ij})+u_{\rm lr}({\bf s}_{ij})\Big)\right],
\label{eq:E3}
\end{equation}
where $q$ is the elementary proton charge, ${\bf s}_{ij}\!=\!{\bf r}_{ij}/\mathfrak{L}$ are scaled (dimensionless) inter-particle
distances, and $\mathfrak{L}$ is the size of the simulation box. In turn, the short- and long-range components of the potential,
and the overall constant $U_{0}$ are given by
\begin{equation}
\begin{split}
  & u_{\rm sr}({\bf s}) = \frac{{\rm erfc}(s/s_{0})}{s}, \\
  & u_{\rm lr}({\bf s}) = \sum_{{\bf l}\neq0} 
   \frac{\exp{(-\pi^{2}s_{0}^{2}l^{2})}}{\pi l^{2}}\exp{(-2\pi i{\bf l}\cdot{\bf s})}, \\
  & U_{0}  =  \frac{Z}{2}u_{\rm lr}(0) - \frac{Z}{\sqrt{\pi}s_{0}} - \frac{\pi s_{0}^{2}Z^{2}}{2},
\label{eq:E3b}
\end{split}
\end{equation}
where ${\bf l}\!=\!(\mathfrak{L}/2\pi){\bf k}$ and $s_{0}$ are the dimensionless momentum and smearing parameter,
respectively, and ${\rm erfc}(s/s_{0})$ is the complementary error function, which is a rapidly decreasing 
function for $s\!>\!s_{0}$. In principle, the Coulomb potential is independent of the value of the smearing 
parameter, but in practice it should neither be too small nor too large, so we fixed it here at 
$s_{0}\!=\!0.12$. For more details see the various appendices on Ref.\cite{R21}.

The cornerstone of the Monte Carlo simulation is the Metropolis algorithm \cite{R30}. Briefly, the $A$ nucleons 
were placed in the simulation box in a random initial configuration, with the initial potential energy denoted by 
$V_{\rm old}({\bf r}_{1},\ldots,{\bf r}_{A})$. Then, each nucleon is moved in succession according to the following 
prescription: ${\bf r}_{i}\rightarrow \xi \Delta{\bf r}$, where $\Delta{\bf r}$ is a constant displacement and 
$\xi\!\in\![-1,1]$ is a uniformly generated random number. The new potential energy 
$V_{\rm new}({\bf r}_{1},\ldots,{\bf r}_{A})$ is then calculated for the new configuration. If the new configuration
is energetically favorable, that is, $V_{\rm new}\!\leq\!V_{\rm old}$, then the move is accepted. If instead 
$V_{\rm new}\!>\!V_{\rm old}$, then the move is accepted with probability 
$\exp\!\big[\!-(V_{\rm new}\!-\!V_{\rm old})/T)\big]$. Note that the kinetic energy of the nucleons is not included
in the energy budget as it is independent of their positions and velocities. The kinetic energy is simply given by
$K\!=\!3AT/2$, where the temperature is introduced to simulate the quantum-mechanical, zero-point motion of the 
nucleons \cite{R15}. To follow the “rule-of-thumb” that the acceptance rate should be in the vicinity of 
50\% \cite{R31}, the displacement $\Delta{\bf r}$ was adjusted until the accepted range fell in the 30-to-50\% 
range. Moreover, we have adopted periodic boundary conditions to compensate for the limited size of the simulation 
box. Finally, we start the MC simulations at an initial temperature of $T\!=\!2\,{\rm MeV}$ where the pasta phases 
are recognizable \cite{R32,R33}, and then cool the system to the final temperature of $T\!=\!1\,{\rm MeV}$, with a
cooling schedule of 0.1 MeV per 4,000 sweeps. We note that the parameters of the model were calibrated at a
temperature of $T\!=\!1\,{\rm MeV}$ \cite{R15}. Once the final temperature is reached, we use 56,000 additional 
sweeps to thermalize the system and finally 1,000 sweeps to accumulate statistics on the physical observables 
of interest. Note that a single sweep consists of $A$ individual MC steps in an attempt to move (on average) one
nucleon per sweep.
\bigskip

\subsection{Pair correlation function and isothermal compressibility}
A fundamental quantity that can be readily extracted from the MC simulations is the pair (or two-body) correlation function 
(PCF). The PCF provides a measure of spatial ordering by representing the conditional probability that a pair of particles
be separated by a given distance. That is,
\begin{equation}
 g_{\mu \nu}(r)\!=\!\frac{\mathfrak{L}^3}{4\pi r^2 N_{\mu} N_{\nu}}
 \left\langle\sum_{i}\sum_{j\neq i}\delta (r-\mid {\bf r}_i-{\bf r}_j \mid)\right\rangle,
\label{eq:E4}
\end{equation}
where $\langle\cdots\rangle$ represents an ensemble average over MC configurations \cite{R34} and $g_{\mu \nu}(r)$ 
is assumed to be normalized to unity as $r$ approached $\mathfrak{L}/2$. Moreover, $\mu$ and $\nu$ represent nucleon indices,
so in total there are three different PCFs of interest. Considering some general features of the PCF and following the analysis
of Ref.\cite{R35}, the PCF was fitted with the following dimensionless functional form with a reduced chi-square 
of $\leq\!0.002$ and a maximum at the contact distance $d$:
\begin{widetext}
\begin{equation}
g(y)= \begin{cases}
   \displaystyle{1+y^{-m}\big[g(d)-1-\lambda\big]\!+\!\left[1+\frac{\lambda-1}{y}\right]e^{-\mathfrak{a} (y-1)}\cos\!{\big[\mathfrak{b}(y-1)\big]}}
   & \text{for } y\geqslant1 , \\
   \displaystyle{g(d)e^{-\theta (y-1)^2}} & \text{for } y<1,
\end{cases}
\end{equation}
\end{widetext}
where $y\!=\!r/d$ and $g(d)$ are the dimensionless inter-particle distance and maximum value of the PCF, respectively \cite{R35}. 
Moreover, $m\!\geqslant\!1$, $\lambda$, $\mathfrak{a}$, $\mathfrak{b}$, and $\theta$ are fitting parameters. This functional form is both accurate and
helps suppress the noise in the simulated PCF due to the limited number of particles. In particular, using the fitted (noise-free) 
PCF one can compute thermodynamic observables that could help identify phase transitions. For example, in a two-component system like the one that we are interested in here, the isothermal compressibility may be written as \cite{R36}:
\begin{equation}
\kappa_T\!=\!\frac{1\!+\!\rho_{\mu}G_{\mu \mu}\!+\!\rho_{\nu}G_{\nu \nu}\!+\!\rho_{\mu}\rho_{\nu}
(G_{\mu \mu} G_{\nu \nu}\!-\!G_{\mu \nu}^2)}{T\Big[\rho_{\mu}+\rho_{\nu}+\rho_{\mu}\rho_{\nu}
(G_{\mu \mu}+G_{\nu \nu}-2G_{\mu \nu})\Big]},
\label{eq:E7}
\end{equation}
where $\rho_{\mu}$ and $\rho_{\nu}$ are the number densities of the two species and $G_{\mu \nu}$ is defined from 
the Kirkwood--Buff theory as the angle-averaged integral of the pair correlation function \cite{R27}:
\begin{equation}
 G_{\mu \nu} =\int_{0}^{\infty} \big[g_{\mu \nu}(r)-1\big]4\pi r^2 dr.
 \label{eq:E8}
\end{equation}

\subsection{Equation of state}
The mass--radius relation of neutron stars in hydrostatic equilibrium, as described by the Tolman--Oppenheimer--Volkoff 
equation, is only sensitive to the equation of state, namely, to the relation between the pressure and the energy density
of the system. The contribution to the energy density of the system from the stellar crust includes the classical kinetic 
energy of the bound clusters, the kinetic energy of the free” neutrons, the kinetic energy of the relativistic electron gas, 
and the expectation value of the potential energy obtained from the MC simulation. Ignoring the contribution from the
non-relativistic heavy clusters, the total energy per nucleon may be written as follows \cite{R21}:
\begin{equation}
\frac{E_{\rm tot}}{A} - m_n = \frac{T_n}{A} + y_p\frac{T_e}{Z} + \frac{\langle V\rangle}{A}.
\label{eq:E10}
\end{equation}
Given that the neutrons bound to the heavy clusters have been eliminated from the above equation, we have designated 
as $T_{n}$ the kinetic energy of the free and bound neutrons. Notably, we used a classical kinetic energy for the bound neutrons, which is the same as protons. Treating the kinetic energy of the electrons and free neutrons as a 
non-interacting Fermi gas, we obtain
\begin{equation}
  T_{\rm FG} \!=\! \frac{3m}{8x_{F}^{3}}\left[x_{F}\sqrt{1\!+\!
  x_{F}^{2}}(1\!+\!2x_{F}^{2})\!-\!\ln{(x_{F}\!+\!\sqrt{1\!+\!x_{F}^{2}})}\right],
 \label{eq:E11}
\end{equation}
where $m$ is the fermion (electron or neutron) mass and $x_{F}$ is the dimensionless Fermi momentum 
\begin{equation}
  x_{F} = \frac{k_{F}}{m} = \frac{(3\pi^{2}\rho)^{1/3}}{m}.
\end{equation}
Although the energy-per-nucleon displayed in Eq.(\ref{eq:E10}) depends on both the overall baryon density and the proton
fraction, what is relevant to the neutron star is the composition in $\beta$-equilibrium, namely, the proton fraction that minimizes 
$E/A$ at a fixed baryon density. However, Eq.(\ref{eq:E10}) remains relevant in connection to a fundamental property of the EoS:
the symmetry energy. Assuming an infinite system of nucleons interacting exclusively via nuclear forces (i.e., no electromagnetic
interactions) the symmetry energy is defined as \cite{R27b}
\begin{equation}
 S(\rho)=\frac{1}{2}\!\left(\frac{\partial^{2}E_{Nucl}/A}{\partial\alpha^{2}}\right)_{\!\!\alpha=0}.
\label{eq:E15}
\end{equation}
where $\alpha\!=\!1\!-\!2y_{p}$ is the neutron-proton asymmetry and $E_{Nucl}$ is the nuclear energy. In particular, in the vicinity of nuclear matter saturation density, the
symmetry energy takes the following simple form \cite{N1}:
\begin{equation}
 S(\rho)=J+L\,\Big(\frac{\rho-\rho_{0}}{3\rho_{0}}\Big),
\label{eq:E16}
\end{equation}
where $J$ is the symmetry energy and $L$ its slope at saturation density. The slope $L$ controls the density dependence of the 
symmetry energy and, in turn, the neutron-proton asymmetry of the bound clusters. For example, a large value of $L$ implies small symmetry energy at the densities of relevance to the stellar crust, thereby allowing for a larger fraction of bound neutrons 
and a corresponding weaker contribution from the free neutrons to the pressure support.

\begin{figure*}[!t]
	\begin{center}
		\includegraphics[trim={6cm 1cm 0 3cm},clip,scale=1,width=16.4cm]{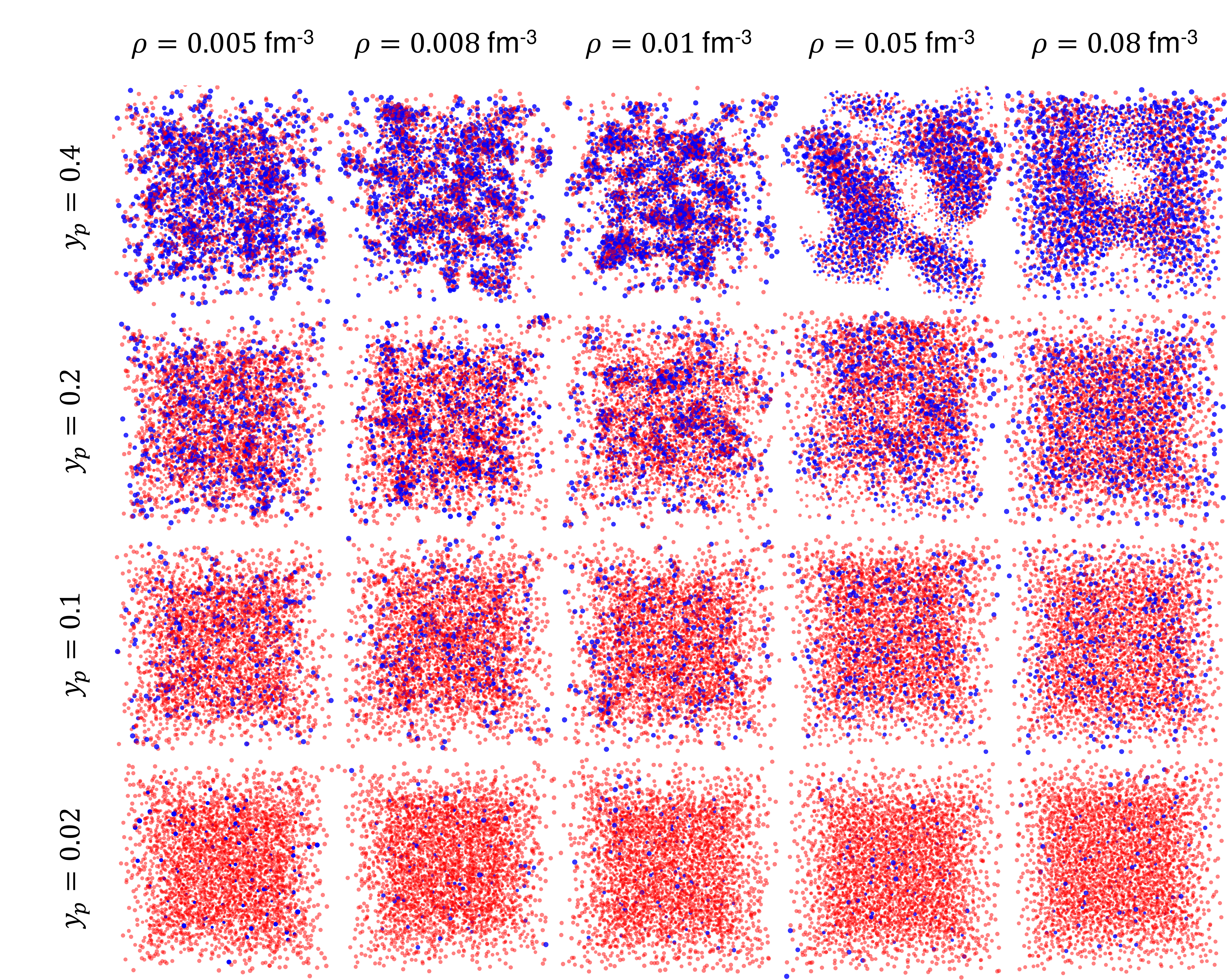}
		\caption{Snapshots of various relaxed (or thermalized) nucleon configurations for a variety of baryon densities
			and proton fractions. To simulate the quantum zero-point motion of the nucleons, all simulations were carried 
			out at a temperature of $T\!=\!1$ MeV. Blue and red solid circles depict protons and neutrons, respectively.} 
		\label{Fig1}
	\end{center}
\end{figure*}

\subsection{Estimation of the bound neutron fraction}

It should be clear that deciding what fraction of the neutrons are bound to clusters and what fraction remains in the vapor is 
a difficult question that is likely to be model-dependent. Nevertheless, we now propose a purely geometrical model to estimate 
the bound neutron fraction $N_{\rm bound}/N$, and consequently the fraction of free neutrons that contribute
to the equation of state. To start, we introduce the concept of the potential of mean force $U(r)$ that is defined in terms of the
proton-neutron correlation function $g_{pn}(r)$\,\cite{N2,N3}. That is,
\begin{equation}
 g_{pn}(r) = \exp\left(-\frac{U_{pn}(r)}{T}\right) \Leftrightarrow U_{pn}(r)=-T\ln g_{pn}(r). 
\label{eq:E16b}
\end{equation}
Although $U_{pn}(r)$ differs from the two-body potential $v_{pn}(r)$ introduced in Eq.(\ref{eq:E2}) due to the presence of the 
medium, the pair potential $v_{pn}(r)$ provides a good approximation to $U_{pn}(r)$ in the dilute limit because of the dominance 
of binary collisions. This suggests that the first peak in the proton-neutron PCF contains critical information on the binding of the 
pair. One can then define a Gaussian function by selecting a length-scale $R_{pn}$ and full width at half maximum $W$ that 
optimally fits the first peak of $g_{pn}(r)$; that is, $g_{pn}(r)$ reaches its maximum value at $R_{pn}$. Finally, we determine that 
a neutron is bound if the distance between such neutron located at a position ${\bf r}_{n}$ and the nearest proton located at  
${\bf r}_{p}$ satisfies:  
\begin{equation}
 \mid {\bf r}_{p}\!-\!{\bf r}_{n} \mid\leqslant R_{pn}\!+\!W.
 \label{eq:E16c}
\end{equation}
To obtain a statistically meaningful quantity, one traces the number of times the position of the given neutron (say neutron $i$) 
satisfies the above condition in all recorded MC configurations. This results in a probability $P_i$ that the neutron is bound.
This procedure is repeated for all $N$ neutrons in the system resulting in a set of independent probabilities $\{P_1,P_2,\ldots,P_N\}$.
This set of probabilities is then used to estimate the number of bound neutrons in the system:
\begin{equation}
\begin{aligned}
 \frac{N_{\rm bound}}{N} = \sum_{i=1}^{N}P_i.
\end{aligned}
\label{eq:E18}
\end{equation}
\vfill

\section{RESULTS AND DISCUSIONS}

\begin{figure*}[ht]
	\begin{center}
		\includegraphics[trim={0 0.25cm 0 0.25cm},clip,scale=1,width=16.2cm]{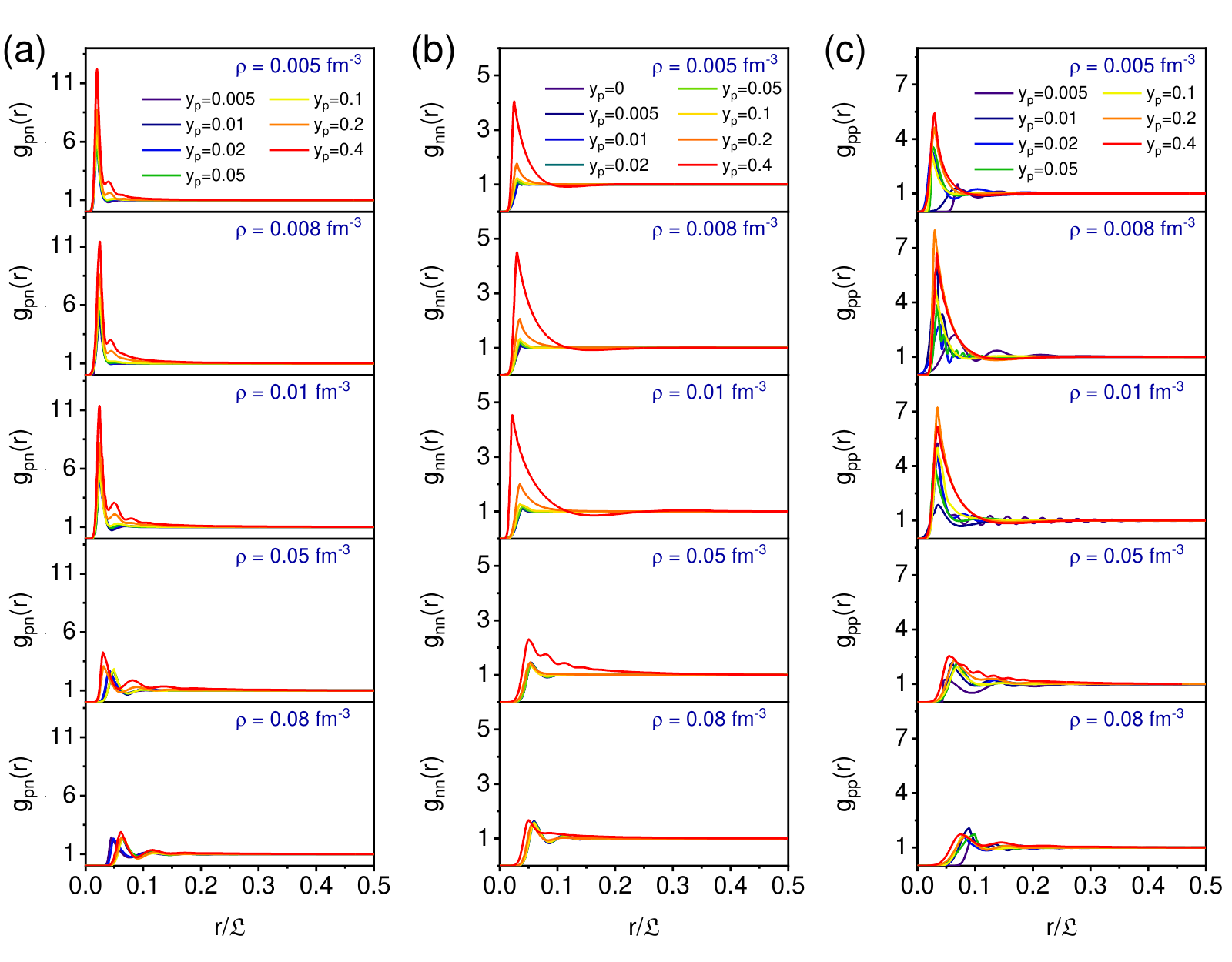}
		\caption{(a) pn, (b) nn, and (c) pp PCFs as a function of the dimensionless 
			distance for different baryon densities and proton fractions.} \label{Fig2}
	\end{center}
\end{figure*}

The main goal of this section is to present results on the various physical observables that can be extracted from the
Monte Carlo simulations. 

\subsection{Relaxed configurations}

We start the section by displaying in Fig.\;\ref{Fig1} snapshots of the relaxed (or thermalized) MC configurations for a wide range 
of baryon densities and proton fractions. At a proton-poor environment of $y_p\!=\!0.02$, protons (Blue solid circles) are well-separated and randomly 
immersed among the neutrons (red solid circles). Given that pure neutron matter does not cluster, the snapshots display a nuclear system that is
nearly uniform. By increasing the proton fraction to the range of $0.1\leq y_p\leq0.2$, the uniformity in the system is broken leading 
to non-symmetric clusters immersed in a uniform neutron vapor. As the proton fraction is increased all the way to $y_p\!=\!0.4$, most
of the neutrons are now absorbed into neutron-rich clusters with perhaps only a hint of a dilute neutron gas; note that $y_p\!=\!0.4$ is
approximately equal to the proton fraction of ${}^{208}$Pb. At a density of $\rho\geq0.01\,{\rm fm}^{-3}$, well defined clusters 
(or nuclei) are observed, which are consistent with earlier reports\,\cite{R15,R38}. Moreover, as the density continues to increase,
one notices the emergence of distinct topological structures, which is a hallmark of “Coulomb frustration”. Under these conditions, 
the typical distance scale is such that it is unclear whether it is energetically favorable for protons to join the clusters in an effort to 
maximize the nuclear attraction or to remain well separated to minimize the Coulomb repulsion. Indeed, at the highest densities of 
$\rho\!=\!0.05$ and $0.08$ fm$^{-3}$, one observe exotic structures such as a waffle-like phase and crossed-cylindrical holes, 
respectively\,\cite{R20}. At slightly higher densities, the uniformity in the system is restored. Given the different density contrasts 
in the pasta phases --- namely, a higher than average density within the clusters --- one would expect a dramatic change in the 
transport properties in the inner crust. To elucidate such behavior, we present the behavior of the isothermal compressibility 
and density fluctuations.

\subsection{Isothermal compressibility}

One of the simplest --- yet fundamental --- physical observables to compute within the Monte Carlo framework is the pair correlation function. 
As noted in Eq.(\ref{eq:E4}), the pair correlation function represents the probability of finding a pair of particles separated by a distance $r$. 
Thus, the PCF may be computed exclusively in terms of the instantaneous positions of the particles. Moreover, the static structure factor $S(q)$, 
obtained from the Fourier transform of the PCF, provides a particularly useful observable that is associated with the mean-square density fluctuations 
of the system. Indeed, the angle-averaged integral of the pair correlation  $G_{\mu \nu}$, defined in Eq.(\ref{eq:E8}) to compute the isothermal
compressibility, is proportional to the static structure factor at zero-momentum transfer. 

In Fig.\,\ref{Fig2} we display PCFs for all three possible nucleon pairs, namely, neutron-neutron (nn), proton-proton (pp), and proton-neutron (pn), for 
different baryon densities and proton fractions. As mentioned earlier, $g_{\mu \nu}$ characterizes the spatial ordering in a many-particle system 
and can be used to calculate the average value of any two-particle physical observable\,\cite{R34}. As expected from the strong nucleon-nucleon repulsion at short distances that produces the hard-core behavior of the NN interaction --- and that is often attributed to the Pauli exclusion principle at the quark level --- $g_{\mu\nu}$ vanishes at small separations.
In the opposite limit of large distances, the PCF goes to one as the probability of finding two nucleons with a given separation is almost constant. 
The height of the first peak is strongly correlated to the number of nearest neighbor nucleons. Given that in our model only the proton-neutron 
interaction is attractive, one expects to find almost all protons within neutron-rich clusters. This is fairly evident in Fig.\,\ref{Fig2}(a), as the size 
of the first peak displayed by the $g_{pn}$ correlation function is significantly larger than both $g_{nn}$ and $g_{pp}$, especially at low densities. 
Moreover, unlike $g_{nn}$ and $g_{pp}$, the proton-neutron PCF display two or three peaks that suggest the presence of a relatively long-ranged 
ordering between protons and neutrons. As depicted in Fig.\,\ref{Fig2}(b), the width of the first peak of $g_{nn}$ remains nearly constant for 
$\rho\!\leq\!0.01\,{\rm fm}^{-3}$, suggesting that the system is dominated by the formation of well-separated individual clusters. Indeed, in the 
dilute regime, it is energetically favorable for nucleons to cluster into “normal” high-density nuclei, which arrange themselves into a Coulomb 
crystal. Instead, for $\rho\!\geq\!0.05\,{\rm fm}^{-3}$ (or about one-third of nuclear saturation density), the height of the first peak decreases 
signaling the evolution towards the uniform phase. The small peak at short distances combined with the oscillatory behavior at larger distances 
illustrates the characteristic behavior of a fluid interacting through short-range forces. Further, as shown in Fig.\,\ref{Fig2}(c), $g_{pp}$ displays 
a similar trend as $g_{nn}$. However, whereas at low density the first peak of $g_{nn}$ remains nearly constant, we found a much steeper rise 
in the size of the first peak of $g_{pp}$ both as a function of density and proton fraction. Also visible is a relatively large shift in the position of 
the first peak value, at low densities and proton fractions, as well as some oscillations at intermediate distances due to the long-range repulsive Coulomb 
interactions. It should be noted that all the above-mentioned statements are consistent with the behavior illustrated in Fig.\,\ref{Fig1} for the relaxed 
configurations. 

\begin{figure}[h]
\begin{center}
\includegraphics[trim={0 0 0 0},clip,scale=1,width=8cm]{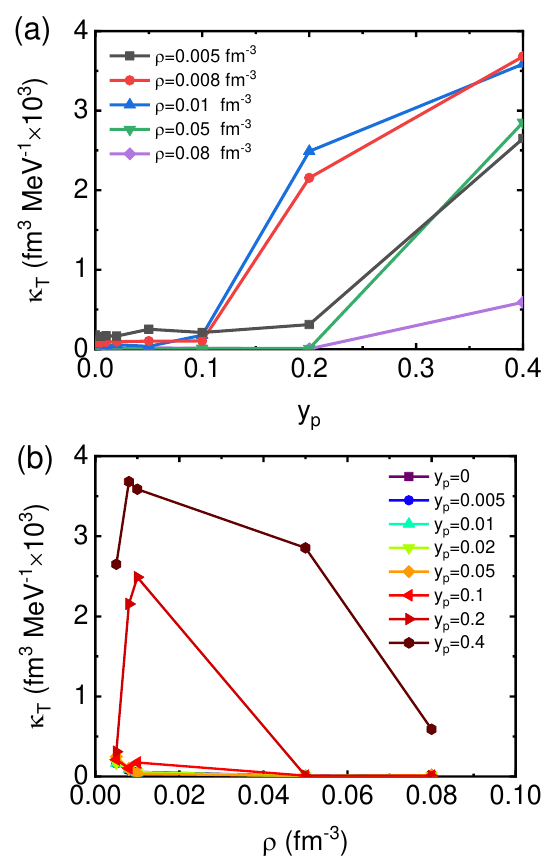}
\caption{(a) Isothermal compressibility as a function of proton fraction for different baryon densities and 
(b) as a function of baryon density for different proton fractions.}
\label{Fig3}
\end{center}
\end{figure}

Although the above qualitative discussions are insightful and informative, one would like to apply a quantitatively robust criterion to identify
the onset of a phase transition. A particularly robust framework in the study of phase transitions is the fluctuation--dissipation theorem, which 
relates fluctuations in the system to the response of the system to a suitable external perturbation. In our particular case, a compelling 
connection involves the isothermal compressibility, the static structure factor, and the fluctuations in the average number of particles. 
Although in the present manuscript we concentrate on the isothermal compressibility as outlined in Eq.(\ref{eq:E7}), in the future, we plan
to simulate a system with a much larger number of particles where the connection between the various indicators of phase transitions will
be made explicit. We display in Fig.\,\ref{Fig3}(a) the isothermal compressibility $\kappa_T$ as a function of the proton fraction for different 
baryon densities; Fig.\,\ref{Fig3}(b) provides the same information but now as a function of the density for a variety of proton fractions.
In the case of proton-poor environments ($y_p\!\lesssim\!0.1$) the thermal compressibility is nearly constant and small, 
suggesting that the system behaves as an incompressible fluid due to the short-range repulsive core of the nn interaction. However, 
$\kappa_T$ shows a striking enhancement of about two orders of magnitude at $y_p\geq0.2$, which is likely associated with the onset 
of a phase transition. The enhancement is driven by the dominance of $G_{pn}\!=\!G_{np}$ over $G_{pp}$ and $G_{nn}$, a fact that
has already been illustrated in Fig.\,\ref{Fig2}. These results show the critical role that the “mixed” pair correlation function $g_{pn}(r)$ 
plays in the thermodynamic properties of the inner crust. To our knowledge, earlier studies have focused exclusively on $g_{pp}$,  
$g_{nn}$, and the associated proton and neutron static structure factors; see Ref.\cite{R21} and references contained therein. 
Ultimately, $\kappa_T$ reaches maximum values of about 2500(3700) at $y_p\!=\!0.2(0.4)$ and $\rho\!=\!0.01(0.008)\,{\rm fm}^{-3}$. 
This behavior is highly reminiscent of a critical point associated with a maximum in the density fluctuations, a phenomenon that has
been extensively studied in the phase diagram of water\,\cite{N4,N5,N6}.

In closing this section we want to underscore the close relationship between various dynamical and thermodynamic quantities that we
plan to simulate in a future work that will involve a much larger number of particles. It is known that for a one-component system of
say $N$ neutrons, the following relations hold:
\begin{equation}
 S(q\!=\!0) = \frac{1}{N}\Big(\langle N^{2} \rangle - \langle N \rangle^{2}\Big) = \frac{NT}{V}\kappa_T\,.
 \label{eq:18b}
\end{equation}

That is, in the limit that the momentum transfer to the system goes to zero (i.e., $q\rightarrow 0$) the static structure factor is related 
to the fluctuations in the number of particles, or equivalently, to the density fluctuations. These fluctuations are themselves related to 
the isothermal compressibility and diverge at the critical point \cite{R38b}. In the particular case of the neutrino opacity, it has been 
argued that the divergence in the density fluctuations near a phase transition may have a dramatic effect on present models of stellar 
collapse\,\cite{R16}.

\subsection{Bound neutron fraction}
\begin{figure}[h]
	\begin{center}
		\includegraphics[trim={0 0 0 0},scale=1,width=8cm]{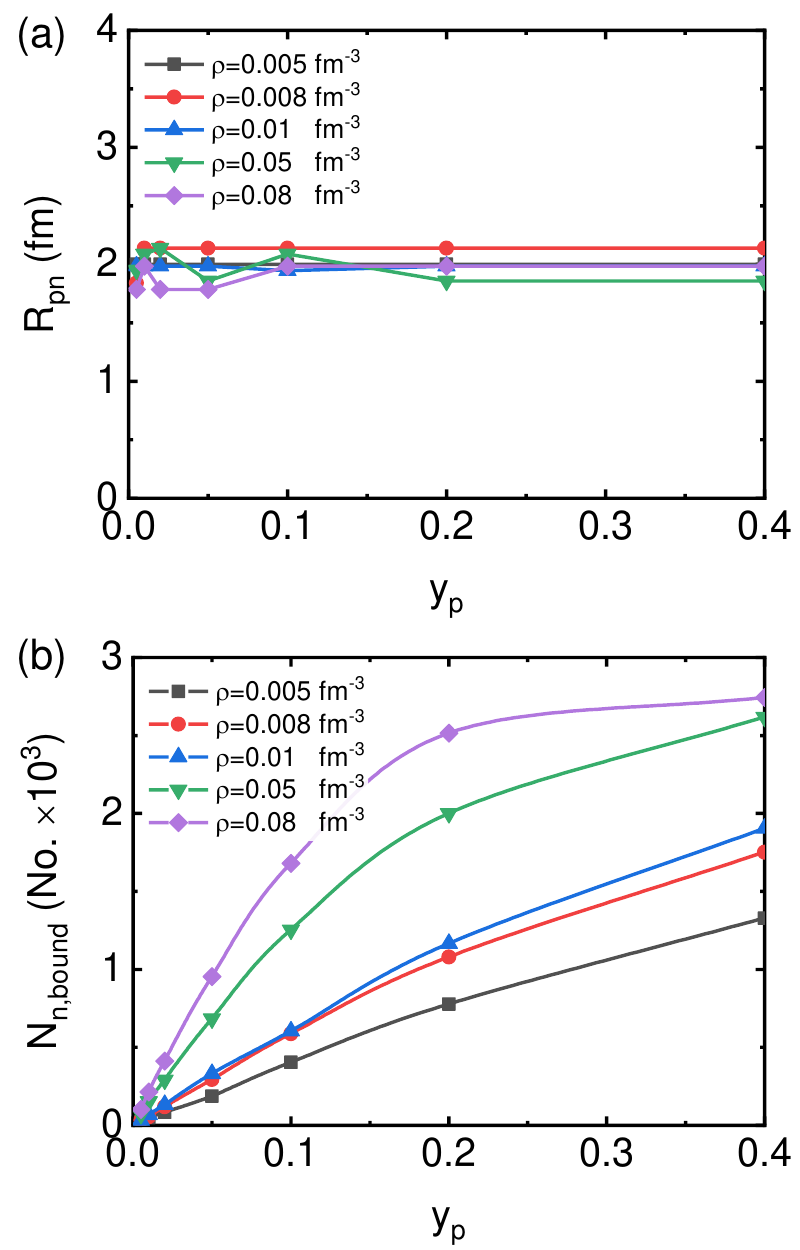}
		\caption{(a) Average proton-neutron distance and (b) number of bound neutrons versus proton fraction. 
			The various lines are added to guide the eye.} \label{Fig4}
	\end{center}
\end{figure}

Calculating the bound neutron fraction not only provides a clear picture of the relaxed configurations but could also have a large 
impact on the total energy of the system. Indeed, the fraction of neutrons not bound into clusters, namely, the “free” neutrons may
provide an efficient source of pressure support in the bottom layers of the inner crust. In this study, we implement the geometrical 
model based on the pn PCF (see Sec.II.D). This geometrical model improves on the simple assumption
adopted in Ref.\cite{R21}. Fig.\;\ref{Fig4}(a) displays the calculated pn average distances $R_{pn}$ as a function of the proton fraction 
for the various densities considered in this work. Our results indicate that the average distance $R_{pn}$ is independent of both $y_p$ 
and $\rho$, giving an average value of $\overline{R}_{pn}\!=\!1.97\pm0.12\,{\rm fm}$. This distance is consistent with the minimum 
value of the two-body pn potential that is located at $1.92\,{\rm fm}$, with the small difference in the central values attributed to 
many-body effects and the small temperature that is included to simulate the quantum zero-point motion. Using such a geometrical 
model, the average number of bound neutrons $N_{\rm bound}$ is displayed in Fig.\,\ref{Fig4}(b). As expected, $N_{\rm bound}$ 
depends strongly on the baryon density and increases monotonically with $y_p$, as adding more protons to the system results in 
the formation of larger neutron-rich clusters.  For example, $N_{\rm bound}$ is significantly higher at $\rho\!=\!0.08\,{\rm fm}^{-3}$ 
than at $\rho\!=\!0.005\,{\rm fm}^{-3}$. As such, the bound neutron fraction can be fitted to the following analytic form:
\begin{equation}
 \frac{N_{\rm bound}}{A}\approx f_{0}\Big(1-e^{-f_{1}\hspace{0.5pt}{y_p}\rho^{f_{2}}}\Big),
 \label{eq:E19}
\end{equation}
where $f_0\!=\!0.6$, $f_{1}\!=\!25.5$, and $f_{2}\!=\!0.52$.
\subsection{Equation of state of the inner crust}
\begin{figure}[tp]
	\begin{center}
		\includegraphics[trim={0 0 0 0},scale=1,width=8cm]{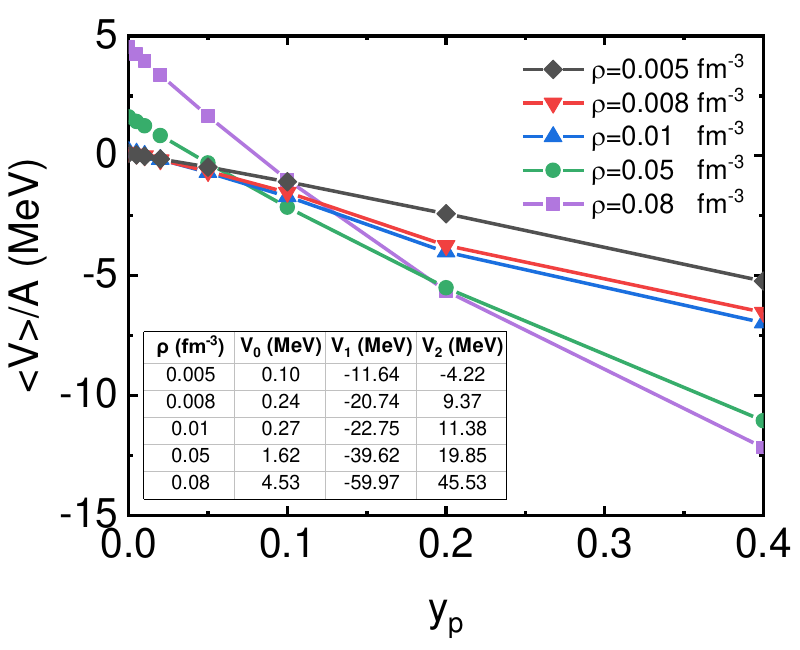}
		\caption{Potential energy per nucleon as a function of the proton fraction for a variety of densities. The inset displays
			the coefficients obtained from a quadratic fit of the form
			$\langle V\rangle/A = V_{0} + V_{1} y_{p} + V_{2} y_{p}^{2}$.}
		\label{Fig5}
	\end{center}
\end{figure}

We conclude this section by displaying in Fig.\,\ref{Fig5} the average potential energy per nucleon $\langle V\rangle/A$ as obtained
from the MC simulations. In agreement with an earlier work by Piekarewicz and Toledo-S\'anchez, \,Ref.\cite{R21}, the potential energy 
is a monotonically decreasing function of $y_p$, given that for small values of the proton fraction the symmetry energy plays a critical 
role. In particular, the $y$-intercept represents the EoS of pure neutron matter as a function of density. At low densities, the inter-particle 
separation exceeds the range of the nn potential so the energy goes to zero. However, as the density reaches $\rho\!=\!0.05\,{\rm fm}^{-3}$, 
the repulsive nn interaction drives the energy to positive values, in accordance with the accepted notion that pure neutron matter is 
uniform and unbound. Further, $\langle V\rangle/A$ declines faster with increasing baryon density because of an increase in the 
bound neutron fraction and, consequently, on the number of attractive pn pairs. For example, for $y_p\!=\!0.4$ and 
$\rho\!=\!0.08\,{\rm fm}^{-3}$, the average potential energy per nucleon is evolving towards the value of the binding energy per 
nucleon of symmetric nuclear matter at saturation density. Note that for each density we have fitted the average potential 
energy per nucleon with a quadratic function of $y_p$ as follows: 
\begin{equation}
 \frac{\langle V\rangle}{A} = V_{0} + V_{1}y_{p} + V_{2}y_{p}^{2} 
 \label{eq:E19b}
\end{equation}
where the appropriate constants are shown in the inset of Fig.\,\ref{Fig5}.
\begin{figure*}[t]
\begin{center}
\includegraphics[trim={0 0 0 0},clip,scale=1,width=16.8cm]{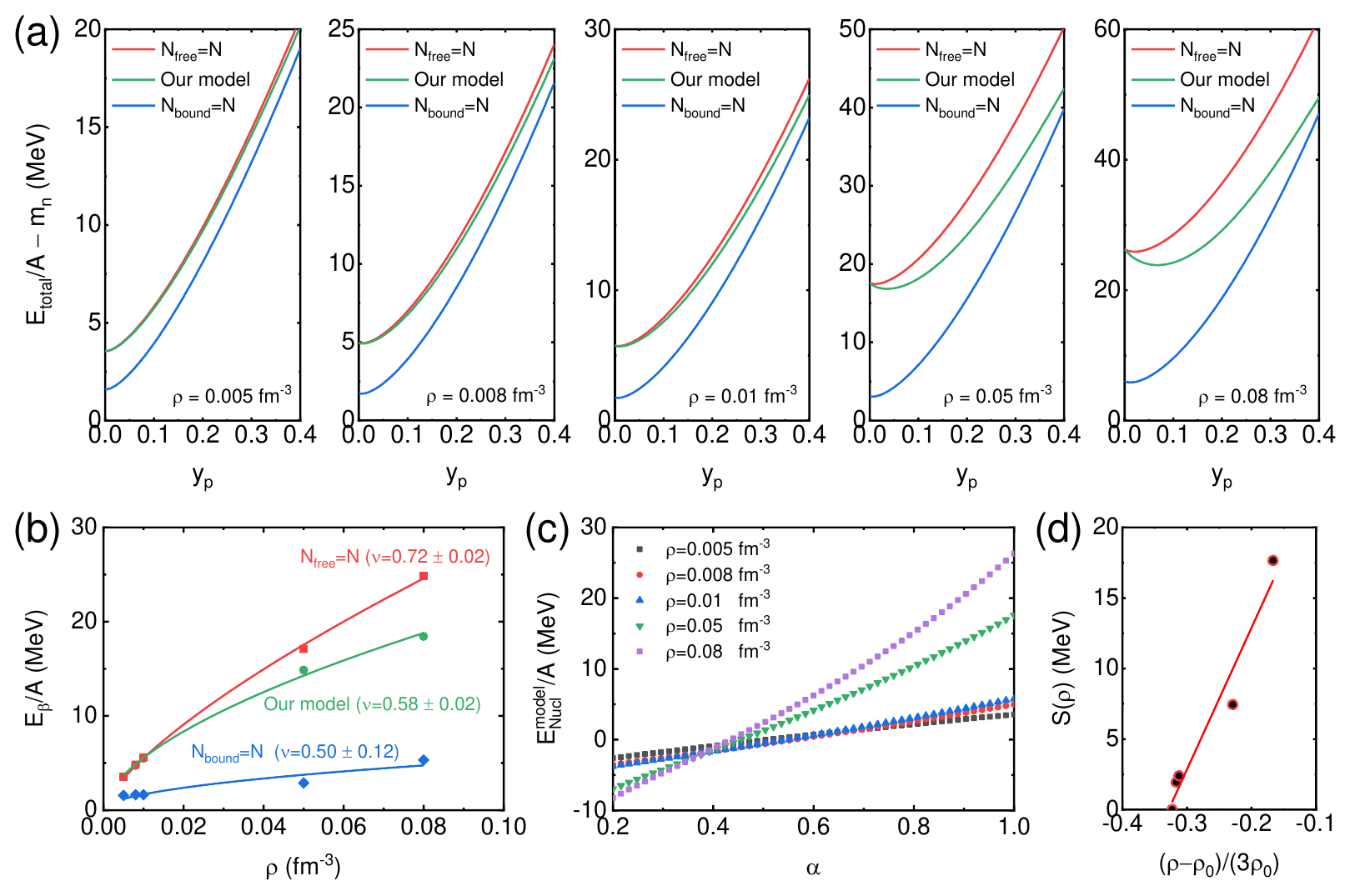}
\caption{(a) Total energy per nucleon versus proton fraction for different baryon densities for our geometrical model (green), 
the quantum limit in which all the neutrons are free (red), and the classical limit in which all neutrons are bound to clusters (blue). 
(b) The corresponding EoS for our model (green), the quantum limit (red), and the classical limit (blue). (c) and (d) Nuclear energy 
per nucleon and symmetry energy versus $\alpha$ and $(\rho\!-\!\rho_{0})/3\rho_{0}$, respectively.} \label{Fig6}
\end{center}
\end{figure*}

Finally, we display in Fig.\,\ref{Fig6}(a) the total energy per nucleon as a function of the proton fraction for different baryon densities and 
different treatments of the neutron contribution to the equation of state; see Eq.(\ref{eq:E10}). Note that the total energy per nucleon 
($E_{\rm tot}/A$) includes the kinetic energy of the electrons. The various panels compare our geometrical model against the quantum 
and classical limits introduced in Ref.\cite{R21}, where either all neutrons are considered as free in the quantum limit or bound in the
classical limits. All curves show a slight initial decline and then an increase with proton fraction, with the former due to the attractive
pn interaction and the latter because of the electronic Fermi energy. For neutron stars in hydrostatic equilibrium, the relevant $T\!=\!0$ 
equation of state is determined from $\beta$ (or chemical) equilibrium. That is, the system settles into the lowest energy configuration 
by finding the optimal value of the proton fraction---a condition that is equivalent to having the $n\!\leftrightarrow p\!+\!e$ in chemical
equilibrium. Although the optimal proton fractions are relatively small, i.e., $0.002\!\leqslant\!y_{p}\!\leqslant\!0.068$ for densities in 
the range of $\rho\!=\!(0.005$--$0.08)\,{\rm fm}^{-3}$, the predicted proton fractions using the geometrical model are still slightly larger 
than the values obtained by assuming that all neutrons are either free or bound\,\cite{R21}. We observe that at very small densities,
the assumption that all neutrons are free is in good agreement with the geometrical model (green lines), but as the density increases, 
it becomes a poor approximation. On the other hand, the assumption that all neutrons are bound into clusters is, as expected, a bad 
approximation at all densities---especially for low proton fractions. We find that our predictions using the geometrical model are
consistent with earlier results using different methods\,\cite{R41,R42,R43,R44,R45}. The nuclear contribution to the total energy
per nucleon in $\beta$-equilibrium is displayed in Fig.\,\ref{Fig6}(b). In particular, for the geometrical model proposed in this work
we predict a nuclear EoS with the power-law behavior $E_{\beta}\!\propto\!\rho^{0.58}$, that is in fairly good agreement with the
EoS of pure neutron matter at low density predicted using quantum Monte Carlo techniques\,\cite{R46}. Given that the solid crust 
accounts for about 10\% of the stellar radius, our results provide a credible estimate of the EoS of the inner edge of the neutron 
star\,\cite{R47,R48}. Finally, we have estimated the nuclear symmetry energy $S(\rho)$ by fitting the nuclear energy as a function 
of the neutron-proton asymmetry $\alpha\!=\!1\!-\!2y_p$, with the results depicted in Fig.\,\ref{Fig6}(c). In turn, the extracted value
of the symmetry energy at the densities considered in this work is displayed in Fig.\,\ref{Fig6}(d). Using a linear fit of the form of Eq.(\ref{eq:E16}).
we obtained the following values for the symmetry energy and its slope at saturation density:
\begin{equation}
\begin{aligned}
 & J=32.96\pm3.68\,{\rm MeV}, \\
 & L=100.34\pm13.29\,{\rm MeV}, \\
\end{aligned}
\label{eq:E21}
\end{equation}
which are comparable to other methods used to extract these values\,\cite{N1,N7,N8}.

\section{CONCLUSIONS}
We have used semi-classical Monte Carlo simulations for a range of baryon densities and proton fractions that
are appropriate for the conditions found in the inner crust of neutron stars. To simulate the quantum zero-point 
motion of the particles, the simulations were carried out at a temperature $T\!=\!1$ MeV with a two-body interaction 
calibrated to reproduce some bulk properties of nuclear matter\,\cite{R15}. Snapshots of the relaxed configurations 
in a proton-poor environment display well-separated clusters immersed in a dilute neutron vapor. Instead, for 
$y_p\geq0.2$, the relaxed configurations display non-symmetric clusters of various topologies that are commonly 
referred to as pasta phases. Further, at the highest simulated proton fraction of $y_p\!=\!0.4$ and densities around $\rho\!=\!0.01\,{\rm fm}^{-3}$, a significant number of “voids” have been identified in the space between 
the isolated, non-symmetric clusters. The appearance of voids results in a highly compressible system as compared 
with other emerging structures. In particular, a dramatic increase in the isothermal compressibility of the two-component 
system was observed at a density of $\rho\!=\!0.008\,{\rm fm}^{-3}$ and a proton fraction of $y_p\!=\!0.4$. To our
knowledge, this is the first time that the Kirkwood--Buff theory has been used to compute the isothermal compressibility 
of the pasta phases. Given the intimate connection between the isothermal compressibility, the long-wave limit of the
static structure factor, and the fluctuations in the number of particles [see Eq.(\ref{eq:18b})], we plan in a future work 
with a much larger number of particles to compute all these three quantities to provide a more robust 
diagnosis of the various phase transitions.  

Besides the use of the Kirkwood--Buff theory for the first time in the characterization of the pasta phases, 
we have also proposed a geometrical model for calculating the bound neutron fraction, which is based on 
the simulation of the pn pair correlation function. Our results have shown that the bound neutron fraction 
increases monotonically with $y_p$  and depends strongly on the overall density of the system. The geometrical 
model improves on the simple prescription adopted in Ref.\cite{R21} that assumed two extreme limits: that either 
all neutrons are bound or all neutrons are free. The present results suggest that none of these limits is valid at 
high densities. Using the geometrical model, we could compute the contribution of the free and bound neutrons 
in the total energy of the system. By following this approach, we were able to extract the equation of state
of cold-catalyzed neutron-star matter in $\beta$-equilibrium. Because of the large electronic contribution to
the energy, $\beta$-equilibrium was reached for small proton fractions, thereby making the contribution from
the free electrons particularly important. In particular, we obtained a parametrized EoS for the inner crust that
could be parametrized as $E_{\beta}\!\propto\!\rho^{0.58}$. Given that the stellar crust contributes by about
10\% of the radius of the neutron star, it would be interesting to find the implications of this finding on the 
mass--radius relation. Finally, although we concentrated exclusively on simulated non-uniform systems, we
extracted some critical parameters of the symmetry energy that are consistent with some recent reports.

\section*{ACKNOWLEDGMENT}
We thank Mohammad Qorbani (from National Taiwan University) for useful discussions, comments, and helping 
to prepare the figures. We acknowledge the High-Performance Computing (HPC) system in the Department of Physics at the University of Tehran, where the MC simulations were performed. One of us (J.P.) acknowledges support from the U.S. Department of Energy Office of Science, Office of Nuclear Physics under Award DE-FG02-92ER40750.

\end{document}